# Moment Generating Functions of Generalized Wireless Fading Channels and Applications in Wireless Communication Theory

Ehab Salahat[(1)], Ali Hakam[(2)], Nazar Ali[(3)], and Ahmed Kulaib[(3)]

[(1)] *Student Member*, IEEE.
[(2)] United Arab Emirates University, Al-Ain, UAE.
[(3)] Khalifa University of Science and Technology, Abu Dhabi, UAE

*Abstract*—In this paper, new exact and approximate moment generating functions (MGF) expression for generalized fading models are derived. Specifically, we consider the $\eta$-$\lambda$-$\mu$, $\alpha$-$\mu$, $\alpha$-$\eta$-$\mu$, $\alpha$-$\lambda$-$\mu$, $\alpha$-$\kappa$-$\mu$, and $\alpha$-$\lambda$-$\eta$-$\mu$ generalized fading distributions to derive approximate MGF expressions. The new expressions are very accurate and, in contrast to earlier results in the literature, avoid any complicated special functions, e.g. the Meijer-$\mathcal{G}$ and Fox $\mathcal{H}$-functions. As such, the new MGF expressions allow easier and more efficient analytical manipulations, which also apply for their special cases such as the Rayleigh, Rice, and Nakagami-*m* fading. As an illustrative application, the average bit error rates for each of the fading models are evaluated using the new derived MGF expressions. The accuracy of the analytical results by using the numerically computed results as a basis of comparison as well as published results from the literature.

## I. Introduction

WIRELESS SYSTEMS are inevitably subjected to different signal propagation impairments, such as the additive Gaussian noise, multipath fading, and shadowing, that predominantly degrade the wireless system performance that is mainly characterized by the bit error rates and ergodic channel capacity [1] [2].

Examples of relatively recent reported generalized wireless fading distributions that gained a considerable attention from the research community include the $\alpha$–$\kappa$–$\mu$ and $\alpha$–$\eta$–$\mu$ models [3] [4]. Those generalized models comprise many other fading models as special cases such as the $\kappa$ – $\mu$ and $\eta$ – $\mu$ [3] [5], were proven to provide better fitting for measurement data as compared to classical fading models [6] [7]. In order to evaluate the performance limits of wireless transmission with the different possible system configurations, moment generating function (MGF) based approaches were suggested to evaluate the error rates [8] [9], ergodic capacity [10], and amount of dispersion [8] to name a few. However, the aforementioned MGF approaches are either dependent on the exact knowledge of the MGF closed-form expression and this is sometimes a very challenging analytical problem, especially for generalized fading models as they are modeled using special functions. Moreover, the MGFs reported in the literature for generalized fading models (e.g. $\alpha$–$\mu$ [11] [12], $\alpha$–$\kappa$–$\mu$ and $\alpha$–$\eta$–$\mu$ [3]) are given in terms of the Fox $\mathcal{H}$- or the Meijer-$\mathcal{G}$ functions, which are computationally inefficient, or alternatively given in terms of their infinite series, and are not friendly for further manipulations. One approach to handle such a problem is to use approximations to the probability density function (PDF) prior to deriving wireless performance expressions. For example, the PDF of the generalized $\alpha$–$\lambda$–$\eta$–$\mu$ was elegantly reduced (estimated) in [13] by recognizing the relation between this model and the $\eta - \mu$ fading model, and is represented as a generalized gamma summands from which the MGF was derived. However, both of the exact and the approximate expressions are given in terms of the $\mathcal{G}$– and $\mathcal{H}$–functions. The other approach, which is adopted in this paper, is to estimate the integrand of the PDF's Laplace transform.

In this paper, the aim is to develop unified MGF expressions for the $\eta$-$\lambda$-$\mu$, $\alpha$-$\mu$, $\alpha$-$\eta$-$\mu$, $\alpha$-$\lambda$-$\mu$, $\alpha$-$\kappa$-$\mu$, and $\alpha$-$\lambda$-$\eta$-$\mu$ generalized models. The expressions are given using simple and efficient mathematical functions, mainly by exploiting the developed exponential approximation of the $exp[-x^r]$ function from [3]. The resulting MGF expressions can be then easily used to derive the $k^{th}$ derivative of the MGF, given as $\mathcal{M}_{end}^{(k)}(s) = (-1)^k \mathcal{E}[\gamma_{end}^k e^{-s\gamma_{end}}]$ where $\gamma$ is the instantaneous signal-to-noise ratio and $\mathcal{E}[\cdot]$ denotes the expectation operation. Such expressions allow simple, direct and unconstrained evaluation of essential performance metrics such as the average bit error rates (for coherent as well as non-coherent digital modulation schemes) and the ergodic capacity.

The rest of the paper is structured as follows. In section II, the considered generalized models are revisited along with the considered exponential approximation. In section III, the new unified MGF expressions are derived for the six generalized wireless fading distributions. Following to that, section IV illustrates some applications by demonstrating the suitability of the new derived MGF expressions to the study of wireless communication systems, where the average bit/symbol error rates are obtained using the derived MGF expressions and are compared with numerically computed ones as well as the results from the literature. Finally, the paper's contributions are summarized in section V.

## II. Preliminaries
### A. The Generalized Fading Models
This section is dedicated to present a brief overview of the six considered generalized fading distribution models, namely the $\alpha$–$\mu$, $\eta$–$\lambda$–$\mu$, $\alpha$–$\kappa$–$\mu$, $\alpha$–$\eta$–$\mu$, $\alpha$–$\lambda$–$\mu$, and $\alpha$–$\lambda$–$\eta$–$\mu$ models. In these models, the fading parameters $\alpha$, $\lambda$, $\eta$, $\mu$ and $\kappa$ account respectively for the nonlinearity, the correlation between the in-phase and quadrature components, the unequal power of these two components, the number of the multipath clusters, and the ratio of the power of the dominant components and the total power of the scattered waves, respectively [4] [6] [7]. In what follows, the symbols $\gamma$ and $\tilde{\gamma}$ denote the instantaneous signal-to-noise-ratio and its average value, respectively.

*i. The $\eta$–$\lambda$–$\mu$ Fading Model*

The PDF of the instantaneous SNR in the $\eta$–$\lambda$–$\mu$ generalized fading model, which was introduced in [14], is given by:

$$f_\gamma(\gamma) = \left[\frac{\sqrt{\pi}\left(2\sqrt{\eta(1-\lambda^2)}\tilde{b}\right)^{2\mu}}{2^{\mu-\frac{1}{2}}\Gamma(\mu)\tilde{d}^{\mu-\frac{1}{2}}\tilde{\gamma}^{\mu+\frac{1}{2}}}\right]\gamma^{\mu-0.5}e^{-\tilde{c}\frac{\gamma}{\tilde{\gamma}}}I_{\mu-\frac{1}{2}}\left(\tilde{d}\frac{\gamma}{\tilde{\gamma}}\right), \quad (1.a)$$

where $I_v(\cdot)$ is the modified Bessel function of the first type [15] [16], and the parameters $\lambda$, $\eta$, $\mu$, $\gamma$ and $\tilde{\gamma}$ are as defined earlier. The PDF (1.a) can be rewritten in a compact form as:

$$f_\gamma(\gamma) = \psi \gamma^{m-1} e^{-\beta\gamma} I_v(d\gamma), \quad (1.b)$$

with $\psi$, $\beta$, $v$, and $m$ are as given in table I, and the internal parameter $\tilde{d} = \tilde{b}\sqrt{(\eta-1)^2 + 4\eta\lambda^2}$, and $\tilde{b}$ and $\tilde{c}$ are given in table II. This model includes the $\lambda$–$\mu$, the $\eta$–$\mu$, the Hoyt, the Rice, the Nakagami-$m$, the Rayleigh, the Exponential, the Gamma, and the One-sided Gaussian models as special cases.

*ii. The $\alpha$–$\mu$ Fading Model*

The PDF of the $\alpha$–$\mu$ generalized fading is given by [11]:

$$f_\gamma(\gamma) = \frac{\alpha\mu^\mu}{2\Gamma(\mu)\tilde{\gamma}^{\alpha\mu/2}}\gamma^{\alpha\mu/2-1}e^{-\left[\frac{\mu}{\tilde{\gamma}^{\alpha/2}}\right]\gamma^{\frac{\alpha}{2}}}$$

$$= \psi\gamma^{m-1}e^{-\beta\gamma^{\tilde{\alpha}}}, \quad (2)$$

with the parameters $\psi$, $m$ and $\tilde{\alpha}$ are as given in table I, and $\gamma$, $\tilde{\gamma}$, $\alpha$ and $\mu$ are as defined earlier. This fading model is also known as the generalized gamma (or Stacy) distribution. This generalized model encompasses many other fading models as particular cases [3] such as the Nakagami-$m$ and the Weibull models. It is worth mentioning that the $\alpha$–$\mu$ model itself is a special case of the generalized models discussed next.

*iii. The $\alpha$–$\eta$–$\mu$, $\alpha$–$\lambda$–$\mu$, $\alpha$–$\kappa$–$\mu$ and $\alpha$–$\lambda$–$\eta$–$\mu$ Fading Models*

These generalized fading models were first introduced in [7] [6] [4]. The models enclose the $\alpha - \mu$, the $\lambda - \mu$, the $\eta - \mu$, the $\kappa - \mu$, the Weibull, the Hoyt, the Rice, the Nakagami-$m$, the Rayleigh, the Lognormal, the Gamma, the Exponential, and the One-sided Gaussian as special cases, by setting the fading parameters to their appropriate values (see [3] [5] for more details). The four generalized models can be written in one general form, which is given by [1]:

$$f_\gamma(\gamma) = \psi\gamma^{m-1}e^{-\beta\gamma^{\tilde{\alpha}}}I_v(d\gamma^{r\tilde{\alpha}}), \quad (3)$$

TABLE I: GENERALIZED FADING MODELS PARAMETERS.

| Model | $\tilde{\alpha}$ | $\psi$ | $\beta$ | $v$ | $d$ | $m$ |
|---|---|---|---|---|---|---|
| $\alpha$–$\mu$ | $\frac{\alpha}{2}$ | $\frac{\tilde{\alpha}\beta^\mu}{\Gamma(\mu)}$ | $\frac{\mu}{\tilde{\gamma}^{\tilde{\alpha}}}$ | – | – | $\tilde{\alpha}\mu$ |
| $\eta$–$\lambda$–$\mu$ | – | $\frac{\sqrt{\pi}\left(2\sqrt{\eta(1-\lambda^2)}\tilde{b}\right)^{2\mu}}{2^{\mu-0.5}\Gamma(\mu)\tilde{d}^{\mu-0.5}\tilde{\gamma}^{\mu+0.5}}$ | $\frac{\tilde{c}}{\tilde{\gamma}}$ | $\mu-\frac{1}{2}$ | $\frac{\tilde{d}}{\tilde{\gamma}}$ | $\mu+\frac{1}{2}$ |
| $\alpha$–$\lambda$–$\eta$–$\mu$ | $\frac{\alpha}{2}$ | $\frac{\tilde{\alpha}\mu^\mu\sqrt{\pi}\left(\mu(1+\eta^{-1})\right)^{m/\alpha}}{(1-\lambda^2)\Gamma(\mu)\tilde{\gamma}^m\tilde{b}^{\mu-0.5}}$ | $\frac{\tilde{c}}{\tilde{\gamma}^{\tilde{\alpha}}}$ | $\mu-\frac{1}{2}$ | $\frac{\mu(1+\eta)\tilde{b}}{2\eta\tilde{\gamma}^{\tilde{\alpha}}}$ | $\tilde{\alpha}\left(\mu+\frac{1}{2}\right)$ |
| $\alpha$–$\lambda$–$\mu$ | $\frac{\alpha}{2}$ | $\frac{2\sqrt{\pi}\tilde{\alpha}h^\mu\mu^{\mu+0.5}}{\Gamma(\mu)H^{\mu-0.5}\tilde{\gamma}^m}$ | $\frac{2h\mu}{\tilde{\gamma}^{\tilde{\alpha}}}$ | $\mu-\frac{1}{2}$ | $\frac{2H\mu}{\tilde{\gamma}^{\tilde{\alpha}}}$ | $\tilde{\alpha}\left(\mu+\frac{1}{2}\right)$ |
| $\alpha$–$\eta$–$\mu$ | $\frac{\alpha}{2}$ | | | | | |
| $\alpha$–$\kappa$–$\mu$ | | $\frac{\tilde{\alpha}\mu(1+\kappa)^{0.5(\mu+1)}}{\kappa^{0.5(\mu-1)}e^{\mu\kappa}\tilde{\gamma}^{0.5\tilde{\alpha}(\mu+1)}}$ | $\frac{\mu(1+\kappa)}{\tilde{\gamma}^{\tilde{\alpha}}}$ | $\mu-1$ | $\frac{2\mu\sqrt{\kappa(1+\kappa)}}{\tilde{\gamma}^{\tilde{\alpha}/2}}$ | $\frac{\tilde{\alpha}(\mu+1)}{2}$ |

TABLE II: INTERNAL PARAMETERS OF THE GENERALIZED MODELS.

| Model | $\tilde{c}$ | $\tilde{b}$ | $h$ | $H$ |
|---|---|---|---|---|
| $\eta$–$\lambda$–$\mu$ | $\frac{\mu(\eta+1)^2}{2\eta(1-\lambda^2)}$ | $\frac{\mu(1+\eta)}{2\eta(1-\lambda^2)}$ | – | – |
| $\alpha$–$\lambda$–$\eta$–$\mu$ | $\frac{\mu(1+\eta)^2}{2\eta(1-\lambda^2)}$ | $\frac{\sqrt{(\eta-1)^2+4\eta\lambda^2}}{1-\lambda^2}$ | – | – |
| $\alpha$–$\lambda$–$\mu$ | – | – | $(1-\lambda^2)^{-1}$ | $\lambda(1-\lambda^2)^{-1}$ |
| $\alpha$–$\eta$–$\mu$ | – | – | $0.25(2+\eta+\eta^{-1})$ | $0.25(\eta^{-1}-\eta)$ |

where $\psi$, $m$, $\beta$, $\tilde{\alpha}$, $v$ and $d$ are given in table I, with the internal parameters $\tilde{c}$, $\tilde{b}$, $h$ and $H$ being defined in table II, where $I_v(\cdot)$ is the modified Bessel function of the first type [15], and $r = 0.5$ for the $\alpha$–$\kappa$–$\mu$ fading and otherwise is equal to 1.

*B. The Exponential Approximation*

In this section, we re-exploit our previously proposed and developed exponential approximation, given in [17] as:

$$e^{-z^{1/\tilde{\alpha}}} \approx \sum_{i=1}^{4} a_i e^{-\ell_i z}, \quad (4)$$

where $a_i$ and $\ell_i$ are fitting parameters as discussed and given in [3], table III. Note that extending (4) to $e^{-sz^{1/\tilde{\alpha}}}$ function, where $s$ is constant, is straightforward.

### III. MOMENT GENERATING FUNCTIONS

The moment generating function (MGF) of the fading PDF can be evaluated as [11]:

$$\mathcal{M}_\gamma(s) = \mathcal{E}[e^{-s\gamma}] = \int_0^\infty f_\gamma(\gamma)e^{-s\gamma}d\gamma, \quad (5)$$

where $\mathcal{E}[\cdot]$ is the expectation (averaging) process. In what follows, we will derive the MGF of each of the generalized fading models described in section II.A.

*i. The MGF of $\eta$–$\lambda$–$\mu$ Fading Model*

Substituting (1) into (5) yields:

$$\mathcal{M}_\gamma(s) = \psi \int_0^\infty \gamma^{m-1} e^{-[\beta+s]\gamma} I_v(d\gamma)d\gamma, \quad (6)$$

which can straightforwardly be evaluated in the closed-form:

$$\mathcal{M}_\gamma(s) = \frac{\psi d^v \Gamma(m+v)}{2^v[s+\beta]^{m+v}\Gamma(v+1)} {}_2F_1\left(\left[\frac{m+v}{2}, \frac{m+v+1}{2}\right]; v+1; \frac{d^2}{[s+\beta]^2}\right), \quad (7).$$

Alternatively, and following a similar approach to that given in [18], one can see that $\int_0^\infty \{\gamma^v I_v(d\gamma)\}e^{-\beta\gamma}d\gamma$ is the Laplace transform of $\gamma^v I_v(\kappa\gamma)\}$, which can be analytically evaluated

$$\mathcal{M}_\gamma(s) \approx \begin{cases} \sum_{i=1}^{4} \frac{a_i d^v \Gamma[m/\tilde{\alpha}+v]\psi}{2^v \tilde{\alpha}[\beta+\ell_i s]^{m/\tilde{\alpha}+v}\Gamma[v+1]} {}_2F_1\left[\left[\frac{m/\tilde{\alpha}+v}{2}, \frac{m/\tilde{\alpha}+v+1}{2}\right], 1+v, \frac{d^2}{[\beta+\ell_i s]^2}\right], & \text{for } r = 1 \\ \sum_{i=1}^{4} \frac{a_i d^v \Gamma[m/\tilde{\alpha}+0.5v]\psi}{2^v \tilde{\alpha}[\beta+\ell_i s]^{[m/\tilde{\alpha}+0.5v]}} {}_1F_1\left[\frac{m}{\tilde{\alpha}}+\frac{v}{2}, v+1, \frac{d^2}{4[\beta+\ell_i s]}\right], & \text{for } r = 0.5 \end{cases} \quad (11)$$

using standard mathematical software (e.g. Mathematica). With some manipulations, (6) can be expressed without the need for any special function in the direct closed-form:

$$\mathcal{M}_\gamma(s) = \left[\frac{4\eta(1-\lambda^2)\tilde{b}^2}{([\tilde{c}+s\tilde{\gamma}]^2 - \tilde{d}^2)}\right]^\mu. \quad (8)$$

Up to our knowledge, the expressions (7) and (8) are new.

*ii. The MGF of α–μ Fading Model*

Following from (5), and substituting (2) with some change of variable, the MGF of the α–μ fading model is given by:

$$\mathcal{M}_\gamma(s) = \frac{\psi}{\tilde{\alpha}} \int_0^\infty \{z^{m/\tilde{\alpha}-1} e^{-\beta z}\} e^{-sz^{1/\tilde{\alpha}}} dz, \quad (9)$$

and by utilizing the estimation (4), one can derive a simple MGF closed-form approximation of the α–μ model, given by:

$$\mathcal{M}_\gamma(s) \approx \sum_{i=1}^{4} \frac{a_i \beta^\mu \Gamma[m/\tilde{\alpha}]}{[\ell_i s+\beta]^{\tilde{m}} \Gamma(\mu)}. \quad (10)$$

Besides the novelty of the expression in (10), evaluating it is very computationally efficient, being a simple sum of scaled Gamma functions. The simplicity of (10) can also be observed by comparing it with the alternative solutions, e.g. eqn. (6) in [11], where the MGF is given by the Meijer-G function with convergence conditions as discussed in therein.

*iii. The MGF of α–η–μ, α–λ–μ, α–λ–η–μ and α–κ–μ Models*

The approximated moment generating functions of the three generalized fading models, namely the α–η–μ, α–λ–μ, α–λ–η–μ models, is in the same format since they have same compact form given in (3), with $r$=1. However, with the α–κ–μ model, the value of $r = 0.5$. By using (3) in (5) and using (4), the unified MGF is found in a simple closed-form as given in (11) for each of the two cases of $r$, at the top of this page. The new MGF expressions in (11) are novel.

## IV. AVERAGE ERROR RATES ANALYSIS

The usefulness of the derived MGF novel expressions (7) – (8) and (10) – (11) was shown by evaluating the average symbol error rates (ASER) using the well-known MGF-approach. For various *M*-ary modulation schemes, such as *M*-ary pulse amplitude modulation (*M*-PAM), *M*-ary phase shift keying (*M*-PSK), *M*-ary differential phase shift keying (*M*-DPSK), as well as the *M*-ary square quadrature amplitude modulation (*M*-QAM) as [19]:

$$P_{SER} = \sum_{\ell=1}^{N} \mathcal{E}_\ell \int_0^{\theta_\ell} \mathcal{M}_{\gamma_{end}}\left(\frac{\phi}{V-2\Lambda\sin^2(\theta)}\right) d\theta, \quad (12)$$

where $N$, $\mathcal{E}_\ell$, $\theta_\ell$, $\phi$, $V$ and $\Lambda$ are given in table III [19].

TABLE III: PARAMETERS SELECTION FOR $P_{SER}$ EVALUATION (12).

| Scheme | N | $\mathcal{E}_\ell$ | $\Lambda$ | $V$ | $\phi$ | $\theta_\ell$ |
|---|---|---|---|---|---|---|
| M-PSK | 1 | $1/\pi$ | -0.5 | 0 | $\sin^2(\pi/M)$ | $\pi(M-1)/M$ |
| M-DPSK | 1 | $2/\pi$ | $\cos(\pi/M)$ | $1+\Lambda$ | $\sin^2(\pi/M)$ | $\pi(M-1)/M$ |
| M-PAM | 1 | $2(1-1/M)/\pi$ | -0.5 | 0 | $3/(M^2-1)$ | $\pi/2$ |
| M-QAM | 2 | $4(1-1/\sqrt{M})/\pi$ | -0.5 | 0 | $1.5/(M-1)$ | $\pi/2$ |
| | | $-4(1-1/\sqrt{M})^2/\pi$ | | | | $\pi/4$ |

For the purpose of comparing and verifying the accuracy of our derived analytical MGF expressions, we use some of the well-known and existing results from the literature as our reference. Specifically, the case of the binary PSK (i.e. *M*-PSK with *M*=2) is assumed here.

Figs. 1 and 2 in [18] are regenerated using (11) and (12). The error rates for different η-λ-μ fading scenarios are also illustrated in Fig. 3 using (7) and (8). By comparing (10) for the α–μ fading model with the numerically obtained results for different scenarios, as shown in Fig. 4. One can clearly see for the different testing conditions, the derived expressions provided accurate plots that closely match the numerically computed ones. These test scenarios confirms the validity of our MGF expressions and their applicability to different fading scenarios.

It can be said again here that the values of the fitting parameters in the approximation (4) can change based on the value of $\alpha$ only. The approximation can be obtained for different values of $\alpha$, where some illustrative values are given in [1][3] for the reader's reference. Experimentally, four exponential terms are sufficient to provide excellent accuracy as shown in this study.

## V. CONCLUSION

In this paper, new exact and approximate MGF expressions for the η-λ-μ, α-μ, α-η-μ, α-λ-μ, α-κ-μ, and α-λ-η-μ generalized wireless fading distributions are derived with the aid of an approximation of the function $exp[-x^r]$. The new expressions are very accurate and, in contrast to earlier results in the literature, avoid any complicated special functions such as the Meijer-$\mathcal{G}$ or the Fox $\mathcal{H}$-function. As such, the new derived MGF expressions allow computationally efficient evaluation and easier analytical manipulation. The error rates for each of the fading models are evaluated to illustrate the applicability and the validity of the new MGF expressions. The accuracy of the derived analytical results was verified numerically and are supported by the existing results from the literature. The authors intend to test and apply these models to indoor environments using real time measurements [20].

## VI. REFERENCES

[1] E. Salahat and A. Hakam, "Novel Unified Expressions for Error Rates and Ergodic Channel Capacity Analysis over Generalized Fading Subject to AWGGN," in *IEEE Global Communications Conference*, Austin, USA, 8-12 Dec. 2014.

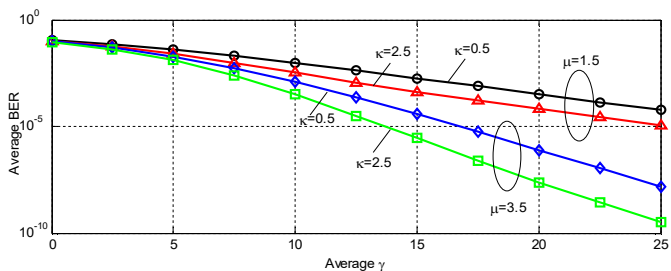

Fig. 1: BPSK error rates for different $\alpha$-$\kappa$-$\mu$ fading scenarios ($\alpha$=2).

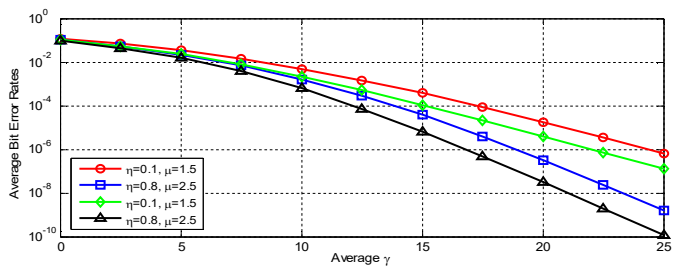

Fig. 2: BPSK error rates for different $\alpha$-$\lambda$-$\eta$-$\mu$ fading scenarios ($\alpha$=2, $\lambda$=0).

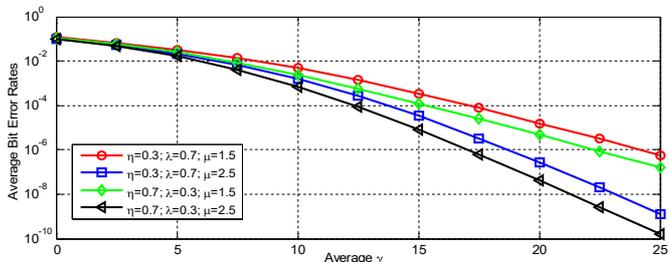

Fig. 3: BPSK error rates for different $\eta$-$\lambda$-$\mu$ fading scenarios.

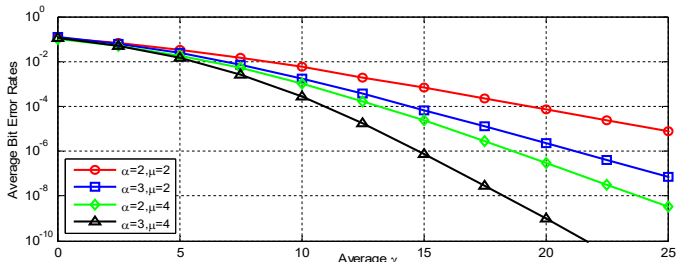

Fig. 4: BPSK error rates for different $\alpha$-$\mu$ fading scenarios.

[2] E. Salahat and I. Abualhoul, "Generalized Average BER Expressions for SC and MRC Receiver over Nakagami-m Fading Channels," in *IEEE International Symposium on Personal, Indoor and Mobile Radio Communications*, London, UK, 8-11 Sept. 2013.

[3] E. Salahat and A. Hakam, "Performance Analysis of Wireless Communications over $\alpha$-$\eta$-$\mu$ and $\alpha$-$\kappa$-$\mu$ Generalized Fading Channels," in *European Wireless Conference (EW'14)*, Barcelona, Spain, 14-16 May 2014.

[4] M. D. Yacoub and G. Fraidenraich, "The $\alpha$-$\eta$-$\mu$ and $\alpha$-$\kappa$-$\mu$ Fading Distributions," in *IEEE International Symposium on Spread Spectrum Techniques and Applications*, Manaus-Amazon, August, 2006.

[5] E. Salahat, "Unified Performance Analysis of Maximal Ratio Combining in $\eta$-$\mu$, $\lambda$-$\mu$ and $\kappa$-$\mu$ Generalized Fading Channels," in *IEEE 80th Vehicular Technology Conference (VTC Fall)*, Vancouver, Canada, 14-17 Sept. 2014.

[6] A. K. Papazafeiropoulos and S. A. Kotsopoulos, "The $\alpha$-$\lambda$-$\eta$-$\mu$: A General Fading Distribution," in *IEEE Wireless Communications and Networking Conference*, Budapest, Hungary, 5-8 Apr. 2009.

[7] A. K. Papazafeiropoulos and S. A. Kotsopoulos, "The $\alpha-\lambda-\mu$ and $\alpha-\eta-\mu$ Small-Scale General Fading Distributions: A Unified Approach," *Wireless Personal Comm.,* vol. 57, no. 4, pp. 735-751, Apr. 2009.

[8] F. Yilmaz and M.-S. Alouini, "Novel MGF-based Expressions for the Average Bit Error Probability of Binary Signalling over Generalized Fading Channels," in *IEEE Wireless Communications and Networking Conference*, Istanbul, 6-9 April 2014.

[9] F. Yilmaz and M.-S. Alouini, "A Novel Unified Expression for the Capacity and Bit Error Probability of Wireless Communication Systems over Generalized Fading Channels," *IEEE Trans. on Communications,* vol. 60, no. 7, pp. 1862 - 1876, 27 June 2012.

[10] K. A. Hamdi, "Capacity of MRC on Correlated Rician Fading Channels," *IEEE Trans. on Communications,* vol. 56, no. 5, p. 708–711, May 2008.

[11] A. M. Magableh and M. M. Matalgah, "Moment Generating Function of the Generalized $\alpha-\mu$ Distribution with Applications," *IEEE Communications Letter,* vol. 13, no. 6, pp. 411-413, June 2009.

[12] D. B. d. Costa and M. D. Yacoub, "Moment Generating Functions of Generalized Fading Distributions and Applications," *IEEE Commun. Lett.,* vol. 12, no. 2, p. 112–114, Feb. 2008.

[13] N. Y. Ermolova and O. Tirkkonen, "Performance Analysis of Communication Systems over Generalized $\alpha-\lambda-\eta-\mu$ Fading Radio Channels," in *IEEE Vehicular Technology Conference (VTC-Spring)*, Dresden, 2-5 June 2013.

[14] A. Papazafeiropoulos and S. Kotsopoulos, "The $\eta$-$\lambda$-$\mu$: A General Fading Distribution," in *IEEE Global Telecommunications Conference*, Honolulu, HI, Nov. 30 2009-Dec. 4 2009.

[15] I. S. Gradshteyn and I. M. Ryzhik, Table of Integrals, Series and Products, San Diego, USA: Academic, 2007.

[16] R. Salahat, E. Salahat, A. Hakam and T. Assaf, "A Simple and Efficient Approximation to the Modified Bessel Functions and Its Applications to Rician Fading," in *IEEE GCC Conference and Exhibition*, Doha, Qatar, 17-20 Nov. 2013.

[17] E. Salahat and H. Saleh, "Novel Average Bit Error Rate Analysis of Generalized Fading Channels Subject to Additive White Generalized Gaussian Noise," in *IEEE Global Conference on Signal and Information Processing*, Atlanta, GA, USA, 3-5 Dec. 2014.

[18] N. Ermolova, "Moment Generating Functions of the Generalized $\eta-\mu$ and $k-\mu$ Distributions and their Applications to Performance Evaluations of Communication Systems," *IEEE Communications Letters,* vol. 12, no. 7, pp. 502-504, Jul. 2008.

[19] F. Yilmaz and O. Kucur, "The M-N Distribution - A New Closed-Form Physical Channel Fading Model and Performance of M-ary Modulations," in *EW'09 Conference*, Aalborg, 17-20 May, 2009.

[20] N. A. Khanbashi, N. A. Sindi, S. Al-Araji, N. Ali, Z. Chaloupka, V. Yenamandra and J. Aweya, "Real Time Evaluation of RF Fingerprints in Wireless LAN Localization Systems," in *10th Workshop on Positioning, Navigation and Comm. (WPNC2013)*, Dresden, 20-21 Mar. 2013.